\documentclass{PoS}

\newcommand{\num}[1]{\color{black} #1}
\graphicspath{ {figs/} }

\title{Four-loop relation between the $\overline{\rm MS}$ and on-shell quark mass}
\ShortTitle{Four-loop relation between the $\overline{\rm MS}$ and on-shell quark mass}

\author{Peter Marquard\\
  Deutsches Elektronen Synchrotron DESY,
  Platanenallee 6, 15738 Zeuthen, Germany\\
  E-mail: \email{peter.marquard@desy.de}}

\author{Alexander V. Smirnov\\
  Scientific Research Computing Center, Moscow State University,
  119991, Moscow, Russia\\
  E-mail: \email{asmirnov80@gmail.com}}

\author{Vladimir A. Smirnov\\
  Skobeltsyn Institute of Nuclear Physics, Moscow State
  University, 119991, Moscow, Russia
  E-mail: \email{smirnov@theory.sinp.msu.ru}}

\author{\speaker{Matthias Steinhauser}
  \\
  Institut f{\"u}r Theoretische Teilchenphysik, Karlsruhe
  Institute of Technology (KIT), 76128 Karlsruhe, Germany\\
  E-mail: \email{matthias.steinhauser@kit.edu}}

\abstract{In this contribution we discuss the four-loop relation
  between the on-shell and $\overline{\rm MS}$ definition of heavy
  quark masses which is applied to the top, bottom and charm case. We also
  present relations between the $\overline{\rm MS}$ quark mass and
  various threshold mass definitions and discuss the uncertainty at
  next-to-next-to-next-to-leading order.}

\FullConference{12th International Symposium on Radiative Corrections
  (Radcor 2015) and LoopFest XIV (Radiative Corrections for the LHC
  and Future Colliders)\\ 15-19 June, 2015\\ UCLA Department of
  Physics \& Astronomy Los Angeles, USA}

\begin{document}


\section{Introduction}

In the Standard Model (and many of its extensions) quark masses enter
as fundamental parameters $m_q$ into the underlying Lagrange density. Once
quantum corrections are considered one has to fix the precise
definition of $m_q$. For heavy quarks a natural one is the on-shell
definition where one requires that the quark propagator $S_q(q)$ has a
pole for $q^2=(M_q^{\rm OS})^2$. However, there are many situations
where other definitions are more convenient. As an example
we mention the decay rate of Higgs bosons to bottom quarks
where, when expressed in terms of the $\overline{\rm MS}$ bottom quark
mass evaluated at the appropriate scale, potentially large logarithms are
automatically summed up. A further example is the threshold production of
top quark pairs in electron-positron annihilation. For this process
one has to adopt a properly constructed (so-called) threshold mass
which, on the one hand, is of short-distance nature as the
$\overline{\rm MS}$ mass. On the other hand, it has similar features 
as the on-shell mass. In particular, it has a physical definition 
at threshold.

It is important to have precise relations among the various 
mass definitions. In Ref.~\cite{Marquard:2015qpa} four-loop
corrections to the relation between the $\overline{\rm MS}$ and
on-shell heavy quark definition has been computed. This result has been
used to derive next-to-next-to-next-to-leading order (N$^3$LO)
relations among the $\overline{\rm MS}$ and the most popular
threshold masses, namely the PS~\cite{Beneke:1998rk},
1S~\cite{Hoang:1998hm,Hoang:1998ng,Hoang:1999zc} and
RS~\cite{Pineda:2001zq} masses.

The relation between the $\overline{\rm MS}$ ($m$) and the on-shell
mass ($M$) is obtained by considering in a first step their relation
to the bare mass, $m_0$:
\begin{eqnarray}
  m^0 = Z_m^{\overline{\rm MS}}m \,,&\quad&
  m^0 = Z_m^{\rm OS} M\,.
  \label{eq::mbare}
\end{eqnarray}
Here, $Z_m^{\overline{\rm MS}}$ is known to five-loop
order~\cite{Baikov:2014qja}. However, in our calculation, only the four-loop
result is necessary~\cite{Chetyrkin:1997dh,Vermaseren:1997fq,Chetyrkin:2004mf}. 
$Z_m^{\rm OS}$ is computed from the scalar and vector contribution of
the quark two-point function with on-shell external momentum via
\begin{eqnarray}
  Z_m^{\rm OS} &=& 1 + \Sigma_V(q^2=M^2) + \Sigma_S(q^2=M^2)
  \,.
\end{eqnarray}
One-, two- and three-loop results to $Z_m^{\rm OS}$ have been computed
in Refs.~\cite{Tarrach:1980up},~\cite{Gray:1990yh}
and~\cite{Chetyrkin:1999ys,Chetyrkin:1999qi,Melnikov:2000qh,Marquard:2007uj},
respectively.  Four-loop result have recently been computed in
Ref.~\cite{Marquard:2015qpa}.

By construction, the ratio of the two equations in~(\ref{eq::mbare})
is finite which leads to 
\begin{eqnarray}
   z_m(\mu) &=& \frac{m(\mu)}{M}
   \,.
  \label{eq::OS2MS}
\end{eqnarray}
It is convenient to cast the perturbative expansion in the form
\begin{eqnarray}
  z_m(\mu) &=& \sum_{n\ge0} \left(\frac{\alpha_s}{\pi}\right)^n z_m^{(n)}
  \,,
  \label{eq::cm}
\end{eqnarray}
with $z_m^{(0)}=1$.
In the next section we present results for $z_m$ up to four-loop order
and discuss the numerical effects for charm, bottom and top quarks.
Afterwards we consider in Section~\ref{sec::thr} the relation between
the $\overline{\rm MS}$ and various threshold masses.
Section~\ref{sec::concl} contains our conclusions.


\section{Four-loop $\overline{\rm MS}$-on-shell relation}

For the computation of the fermion self energy we
use an automated setup which generates all contributing amplitudes
with the help of {\tt qgraf}~\cite{Nogueira:1991ex}. The output is transformed to
{\tt FORM3}-readable~\cite{Vermaseren:2000nd} input using {\tt q2e} and {\tt
  exp}~\cite{Harlander:1997zb,Seidensticker:1999bb}. Afterwards projectors for
the scalar and vector part are applied, traces are taken and the
scalar products in the numerator are decomposed in propagator factors.
This leads to several million different integrals encoded in functions with 
14 different indices which belong to 100 different integral families. 

The Laporta algorithm~\cite{Laporta:2001dd} is applied to each family using
{\tt FIRE5}~\cite{Smirnov:2014hma} and {\tt crusher}~\cite{crusher} 
which are written in {\tt C++}.  Then we use
the code {\tt tsort}~\cite{Pak:2011xt}, which is part of the latest {\tt FIRE}
version, to reveal relations between primary master integrals (following
recipes of Ref.~\cite{Smirnov:2013dia}) and end up with 386 four-loop massive
on-shell propagator integrals, i.e. with $p^2=M^2$.

Up to this point the whole calculation is analytic. However, at the
moment not all master integrals could be evaluated analytically
but only numerically using {\tt
  FIESTA}~\cite{Smirnov:2008py,Smirnov:2009pb,Smirnov:2013eza}
which leads to an accuracy of about five to six digits for the highest
$\epsilon$ expansion term. For some integrals a two- or threefold
Mellin Barnes representation could be derived which enabled us to
obtain a precision of more than eight, in some cases even more
than 20 digits.

For each integral which is evaluated numerically, each $\epsilon$
coefficient gets a separate uncertainty assigned.  Since it results
from a numerical Monte Carlo integration we interpret it as a standard
deviation and combine the individual uncertainties in the final
expression quadratically. Furthermore, we multiply the uncertainty in
the final result for the $\overline{\rm MS}$ and on-shell relation by
a factor five.

Note that we have performed the calculation allowing for a general gauge
parameter $\xi$ keeping terms up to order $\xi^2$ in the expression we
give to the reduction routines. We have checked that $\xi$ drops out
after mass renormalization but before inserting the master integrals.

In the following we show the $\overline{\rm MS}$-on-shell relation in
the form where the on-shell mass is computed from the $\overline{\rm
  MS}$ mass. We discuss the top, bottom and charm quark case and use
as input the following $\overline{\rm MS}$ masses:
$m_t\equiv m_t(m_t) = 163.643$~GeV, 
$m_b\equiv m_b(m_b) = 4.163$~GeV~\cite{Chetyrkin:2009fv},
and
$m_c(3~GeV) = 0.986$~GeV~\cite{Chetyrkin:2009fv}.
The corresponding values for the strong coupling are given by
$\alpha_s^{(6)}(m_t)=0.1088$,
$\alpha_s^{(5)}(m_b)=0.2268$, and
$\alpha_s^{(4)}(3~\mbox{GeV})=0.2560$.
They have been computed from
$\alpha_s^{(5)}(M_Z)=0.1185$~\cite{Agashe:2014kda} using 
{\tt RunDec}~\cite{Chetyrkin:2000yt,Schmidt:2012az}.
In the case of the charm quark we also provide results for
$\mu=m_c$ using the input values
$m_c\equiv m_c(m_c) = 1.279$~GeV and 
$\alpha_s^{(4)}(m_c)=0.3923$.
Note that the choice $\mu=3$~GeV is preferable since
it has the advantage that low renormalization scales $\mu\approx m_c$
are avoided. Our results read
\begin{eqnarray}
  M_t &=& m_t\left(
    1 + 0.4244 \,{\alpha_s} + 0.8345 \,{\alpha_s^2} + 2.375 \,{\alpha_s^3}
    + (8.49 \pm 0.25) \,{\alpha_s^4}
  \right)
  \nonumber\\&=&\mbox{}
  163.643 + 7.557 + 1.617 + 0.501
  + 0.195 \pm 0.005~\mbox{GeV}
  \,,
  \label{eq::mt}
\end{eqnarray}
\begin{eqnarray}
  M_b &=& m_b\left(
    1 + 0.4244 \,{\alpha_s} + 0.9401 \,{\alpha_s^2} + 3.045 \,{\alpha_s^3}
    + (12.57 \pm 0.38) \,{\alpha_s^4}
  \right)
  \nonumber\\&=&\mbox{}
  4.163 + 0.401 + 0.201 + 0.148
  + 0.138 \pm 0.004~\mbox{GeV}
  \,,
  \label{eq::mb}
\end{eqnarray}
\begin{eqnarray}
  M_c &=& m_c(3~\mbox{GeV})\left(
    1 + 1.133 \,{\alpha_s} + 3.119 \,{\alpha_s^2} + 10.98 \,{\alpha_s^3}
    + (51.29 \pm 0.52) \,{\alpha_s^4}
  \right)
  \nonumber\\&=&\mbox{}
  0.986 + 0.286 + 0.202 + 0.182
  + 0.217 \pm 0.002~\mbox{GeV}
  \,,
  \label{eq::mc3}
\end{eqnarray}
\begin{eqnarray}
  M_c &=& m_c\left(
    1 + 0.4244 \,{\alpha_s} + 1.0456 \,{\alpha_s^2} + 3.757 \,{\alpha_s^3}
    + (17.36 \pm 0.52) \,{\alpha_s^4}
  \right)
  \nonumber\\&=&\mbox{}
  1.279 + 0.213 + 0.206 + 0.290
  + 0.526 \pm 0.016~\mbox{GeV}
  \,.
  \label{eq::mc}
\end{eqnarray}
For the top quark the higher order corrections become successively smaller by
a factor two to three leading to a four-loop correction term of about
200~MeV. This is the same order of magnitude as the intrinsic uncertainty of
the $\overline{\rm MS}$-on-shell relation given by $\Lambda_{\rm QCD}$. The
four-loop corrections are still smaller than the current uncertainty of the
top quark form the TEVATRON and the
LHC~\cite{ATLAS:2014wva}.  However, they are not negligible.

For the bottom and charm quark case the situation is completely
different. No convergence is observed when increasing the loop order.
In the case of the charm quark where $m_c(m_c)$ is chosen as a
starting point one even observes a four-loop coefficient which is
almost twice as large as the three-loop one.

From the above results one can conclude that the immediate application
of the $\overline{\rm MS}$-on-shell relation is only meaningful for
the top quark case. For the lighter quarks the on-shell mass parameter
should be avoided. If necessary an appropriately chosen threshold mass
should be used as we will discuss in the next section.

In the following we present for the top quark mass the 
inverted relation of Eq.~(\ref{eq::mt}) 
which reads\footnote{Note that the $\overline{\rm MS}$ value
  used in Eq.~(\ref{eq::mt}) has been obtained using 
  Eq.~(\ref{eq::mtSI}) to three-loop accuracy.}
\begin{eqnarray}
  m_t &=& M_t\left(
    1 - 0.4244 \,{\alpha_s} - 0.65441 \,{\alpha_s^2} - 1.944 \,{\alpha_s^3}
    - (7.23 \pm 0.22) \,{\alpha_s^4}
  \right)
  \nonumber\\&=&\mbox{}
 173.34 - 7.948 - 1.324 - 0.425
  - 0.171 \pm 0.005~\mbox{GeV}
  \,,
  \label{eq::mtSI}
\end{eqnarray}
where $M_t=173.34$~GeV~\cite{ATLAS:2014wva} and
$\alpha_s^{(6)}(M_t)=0.1080$ has been used.
This equation can be used to compute
$m_t(m_t)$ for a given value for the on-shell mass $M_t$.


\section{\label{sec::thr}Relation between $\overline{\rm MS}$ and
  threshold masses to N$^3$LO}  

In this section we present numerical results for the $\overline{\rm
  MS}$ quark masses using input values for the PS, 1S and RS threshold
masses. In practical applications the latter are extracted from
comparisons with experimental data.  The derivation of the N$^3$LO
relations is discussed in Ref.~\cite{Marquard:2015qpa} following the
prescriptions provided in the original
references~\cite{Beneke:1998rk,Hoang:1998hm,Hoang:1998ng,Hoang:1999zc,Pineda:2001zq}. 

Table~\ref{tab::mtMS} shows results for the 
$\overline{\rm MS}$ top quark mass computed from the PS, 1S and RS 
threshold mass values given in the first and second row.
Note that these values are chosen in such a way that in all three
cases the same $\overline{\rm MS}$ mass is obtained after 
applying four-loop corrections, which facilitates the comparison. 
Note also, that in contrast to the corresponding table in
Ref.~\cite{Marquard:2015qpa} we choose for the factorization scale of
the PS mass $\mu_f=80$~GeV instead of $\mu_f=20$~GeV. This is
suggested by the N$^3$LO threshold analysis of $\sigma(e^+e^-\to
t\bar{t})$ performed in Ref.~\cite{Beneke:2015kwa}.
The factorization scale for the RS mass is kept at $\mu_f=20$~GeV.

\begin{table}[ht]
\begin{center}
\begin{tabular}{c|ccc}
input & $m^{\rm PS} =$ & $m^{\rm 1S} = $ & $m^{\rm RS} =$ \\
\#loops  &    168.204 &    172.227 &    171.215\\
\hline
1       &    164.311    &    165.045    &    164.847 \\
2       &    163.713    &    163.861    &    163.853 \\
3       &    163.625    &    163.651    &    163.663 \\
4       &    163.643    &    163.643    &    163.643 \\
\hline
4 ($\times 1.03$)       &    163.637    &    163.637    &    163.637
\\
\end{tabular}
\caption{\label{tab::mtMS}$m_t(m_t)$ in GeV computed from the PS, 1S
  and RS quark mass using LO to N$^3$LO accuracy. The numbers in
  the last line are obtained by taking into account the uncertainty of
  the four-loop coefficient, i.e., it is increased by {\num 3\%}.}
\end{center}
\end{table}

In all three cases one observes a rapid convergence of the perturbative
series. In fact, the NNLO term amounts to at most 210~MeV (1S mass), and at
N$^3$LO at most 20~MeV (RS mass).  After increasing the four-loop
$\overline{\rm MS}$-on-shell term by 3\%, which is the current uncertainty on
the four-loop coefficient in Eq.~(\ref{eq::cm}), the mass values reduces by
6~MeV. Combining these two sources of uncertainties one ends up in a final
uncertainty below 20~MeV which is sufficient for a precise determination of
$m_t$ at a future linear collider~\cite{Beneke:2015kwa}.

\begin{figure}[tb]
  \centering
  \includegraphics[width=\linewidth]{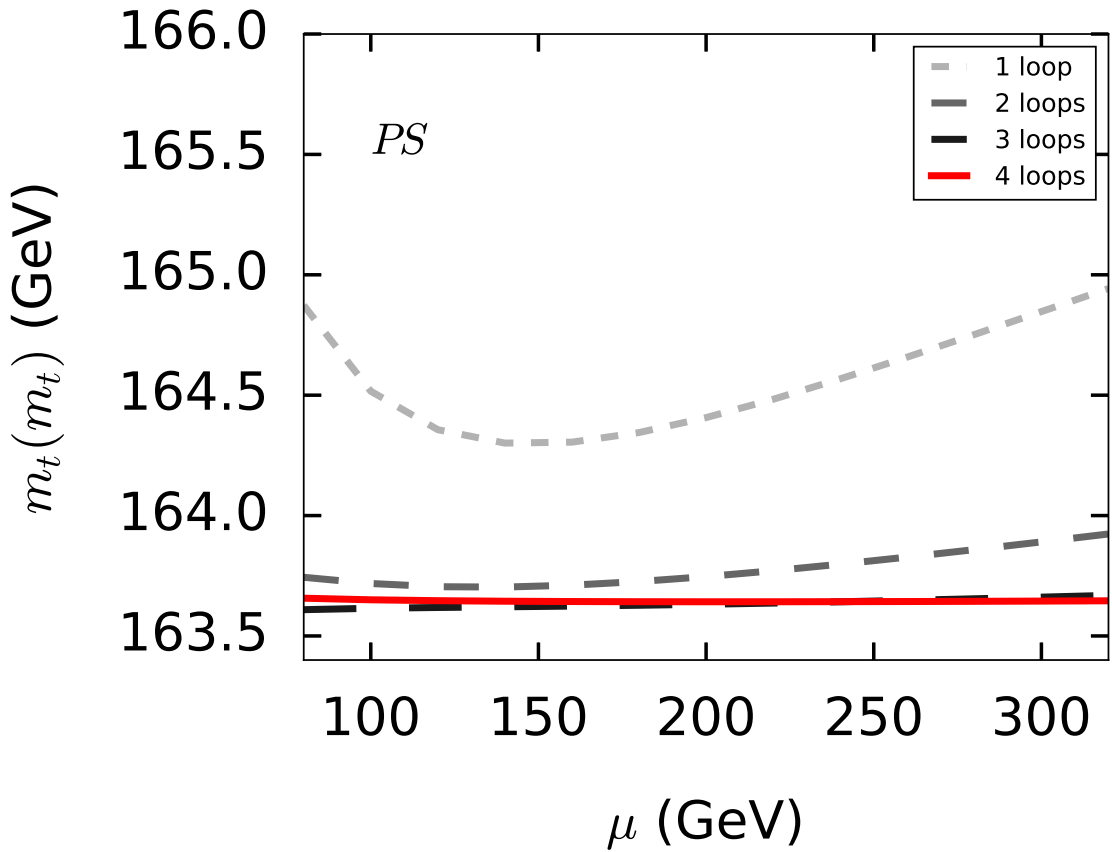}
  \caption{$\overline{\rm MS}$ top quark mass $m_t(m_t)$ 
  computed from the PS mass with LO, NLO, NNLO and N$^3$LO accuracy
  as a function of the renormalization scale used in the 
  $\overline{\rm MS}$-threshold mass relation.}
  \label{fig::mtmt_PS}
\end{figure}

Let us at this point have a closer look to the PS mass.
In Table~\ref{tab::mtMS} the renormalization scale has been fixed to
$\mu=m_t$. It is also interesting to consider different values of
$\mu$ and compute in a first step $m_t(\mu)$ which is then evolved to
$m_t(m_t)$ using renormalization group methods. In
Figure~\ref{fig::mtmt_PS} we plot the result for $m_t(m_t)$
computed from $m_t^{\rm PS}=168.204$~GeV using LO, NLO, NNLO and
N$^3$LO accuracy (from short-dashed to solid lines). Whereas the LO
curve shows a variation of several hundred MeV the 
N$^3$LO is basically independent of $\mu$. Actually, in the considered
range from $m_t/2$ to $2 m_t$ it varies by less than 20~MeV, 
a number comparable to the difference between the NNLO
and N$^3$LO result at the central scale $\mu=m_t$.

\begin{figure}[tb]
  \centering
  \includegraphics[width=\linewidth]{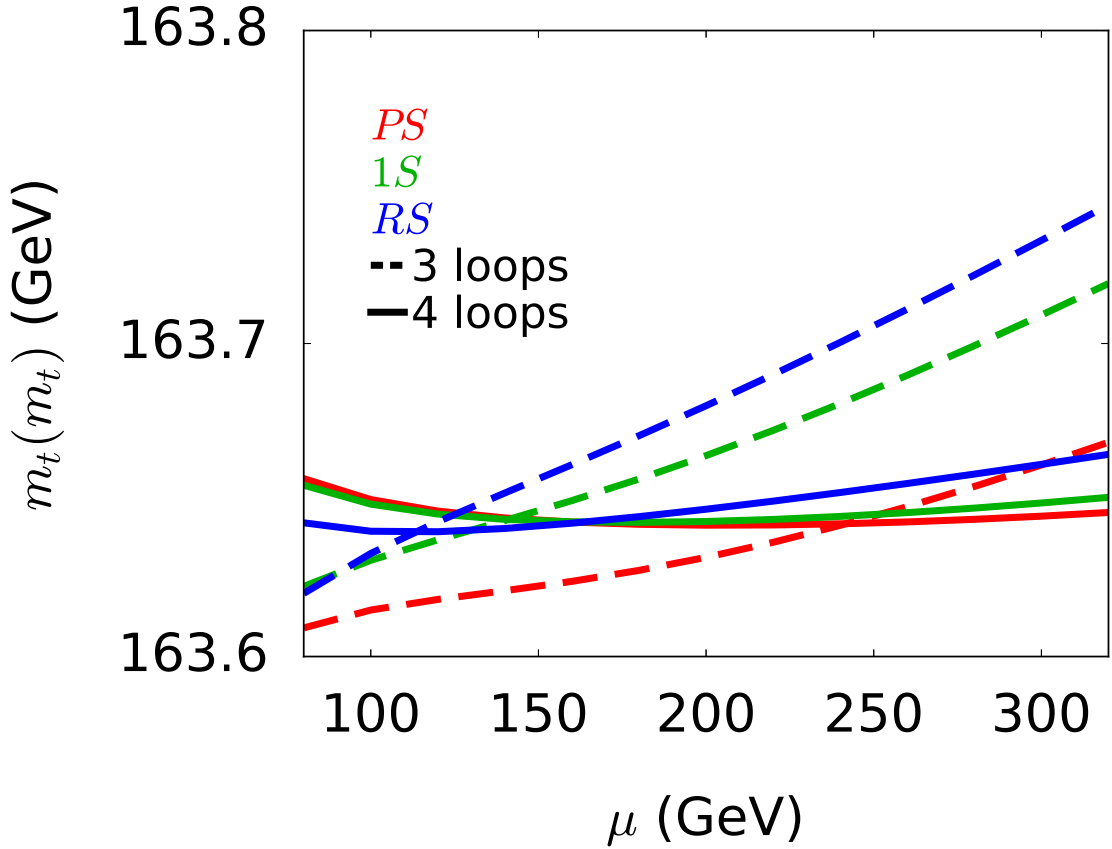}
  \caption{$\overline{\rm MS}$ top quark mass $m_t(m_t)$ computed from
    the PS, 1S and RS mass with NNLO (dashed) and N$^3$LO (solid line)
    accuracy as a function of the renormalization scale used in the
    $\overline{\rm MS}$-threshold mass relation. For $\mu=300$~GeV
    the lines from bottom to top correspond to the PS, 1S and RS mass.}
  \label{fig::mtmt}
\end{figure}

The behaviour of the NNLO and N$^3$LO curve of
Figure~\ref{fig::mtmt_PS} is magnified in Figure~\ref{fig::mtmt}
(red curves). In addition the corresponding results are shown 
for the 1S (green) and RS (blue) mass. In all three cases one observes
a significant improvement of the $\mu$ dependence when going from
NNLO to N$^3$LO. Furthermore, the N$^3$LO curves of 
all three threshold masses only depend mildly on $\mu$.

In Table~\ref{tab::mbMS} results for the $\overline{\rm MS}$ bottom
quark mass are shown. They are computed from the PS, 1S and RS masses
as given in the first and second row of the table using LO, NLO, NNLO
and N$^3$LO accuracy. Similar to the top quark case, one observes a
rapid convergence with a shift below 10~MeV from the last perturbative
order. A variation of the four-loop $\overline{\rm MS}$-on-shell
coefficient leads to a shift of 4~MeV in the $\overline{\rm MS}$ mass.

\begin{table}[t]
\begin{center}
\begin{tabular}{c|ccc}
input & $m^{\rm PS} =$ & $m^{\rm 1S} = $ & $m^{\rm RS} =$ \\
\#loops  &      4.483 &      4.670 &      4.365\\
\hline
1       &      4.266    &      4.308    &      4.210 \\
2       &      4.191    &      4.190    &      4.172 \\
3       &      4.161    &      4.154    &      4.158 \\
4       &      4.163    &      4.163    &      4.163 \\
\hline
4 ($\times 1.03$)       &      4.159    &      4.159    &      4.159 \\
\end{tabular}
\caption{\label{tab::mbMS}$m_b(m_b)$ in GeV computed from the PS, 1S
  and RS quark mass using LO to N$^3$LO accuracy. The numbers in
  the last line are obtained by taking into account the uncertainty of
  the four-loop coefficient, i.e., it is increased by {\num 3\%}.
  The factorization scales for the PS and RS mass are set to 2~GeV.}
\end{center}
\end{table}

In Table~\ref{tab::mc3MS} we show the corresponding results for the
$\overline{\rm MS}$ charm quark mass. Even in this case we observe a
reasonable convergence of the perturbative series. For the PS and RS
mass the N$^3$LO corrections are even below 10~MeV.

\begin{table}[t]
\begin{center}
\begin{tabular}{c|ccc}
input & $m^{\rm PS} =$ & $m^{\rm 1S} = $ & $m^{\rm RS} =$ \\
\#loops  &      1.155 &      1.552 &      1.044\\
\hline
1       &      1.078    &      1.265    &      1.028 \\
2       &      1.021    &      1.119    &      1.008 \\
3       &      0.993    &      1.033    &      0.991 \\
4       &      0.986    &      0.986    &      0.986 \\
\hline
4 ($\times 1.03$)       &      0.984    &      0.984    &      0.984
\\
\end{tabular}
\caption{\label{tab::mc3MS}$m_c(3~\mbox{GeV})$ in GeV computed from the PS, 1S
  and RS quark mass using LO to N$^3$LO accuracy. The numbers in
  the last line are obtained by taking into account the uncertainty of
  the four-loop coefficient, i.e., it is increased by {\num 3\%}.
  The factorization scales for the PS and RS mass are set to 2~GeV.}
\end{center}
\end{table}


\section{\label{sec::concl}Conclusions}

In this contribution we considered the four-loop relation between the
$\overline{\rm MS}$ and on-shell heavy quark masses and applied it to
the top, bottom and charm case. Whereas the perturbative series
converges well for top it does not for the other two cases. This
suggests that the on-shell top quark mass is a reasonably good
parameter at the order of 100~MeV or even better.  For all three cases
the perturbative relation between the threshold (PS, 1S, RS) and the
$\overline{\rm MS}$ masses is perturbatively well behaved. Thus, in
case a threshold mass is determined from a physical quantity like a
(threshold) cross section or a bound state energy it can be related to
the corresponding $\overline{\rm MS}$ mass with high precision.


\section*{Acknowledgments}

We thank the High Performance Computing Center
Stuttgart (HLRS) and the Supercomputing Center of Lomonosov Moscow
State University~\cite{LMSU} for providing computing time used for the
numerical computations with {\tt FIESTA}.  P.M. was supported in part
by the EU Network HIGGSTOOLS PITN-GA-2012-316704.  This work was
supported by the DFG through the SFB/TR~9 ``Computational Particle
Physics''.  The work of V.S. was supported by the Alexander von
Humboldt Foundation (Humboldt Forschungspreis).


\end{document}